\documentclass[twocolumn,showpacs,preprintnumbers,amsmath,amssymb,floatfix]{revtex4}

\usepackage{graphicx}
\usepackage{dcolumn}
\usepackage{bm}
\begin{document}


\title{Coarse-Grained Back Reaction in Single Scalar Field Driven Inflation}

\author{Ghazal Geshnizjani}

\affiliation{Department of Physics, Brown University, Providence,
RI 02912, USA}
\email{ghazal@het.brown.edu}

\author{Niayesh Afshordi}
\affiliation{Princeton University Observatory, Princeton
University, Princeton, NJ 08544, USA}

\begin{abstract}
We introduce a self-consistent stochastic coarse-graining method,
which includes both metric and scalar field fluctuations, to
investigate the back reaction of long wavelength perturbations in
single-scalar driven inflation, up to the second (one loop) order.
We demonstrate that, although back reaction cannot be significant
during the last 70 e-foldings of inflation with a smooth
potential, there exist non-smooth inflaton potentials which allow
significant back reaction, and are also consistent with
cosmological observations. Such non-smooth potentials may lead to
the generation of massive primordial black holes, which could be
further used to constrain/verify these models.
\end{abstract}


\maketitle

\section{Introduction}
In the context of inflationary cosmology, initial seeds of today's
structures are generated in the de-Sitter phase of an inflationary
universe, as a consequence of quantum vacuum fluctuations (see
e.g. \cite{MFB} for a comprehensive overview of the theory of
cosmological fluctuations and \cite{B03} for a recent introductory
overview). The standard way of analyzing these fluctuations is
through linear perturbation theory. However, the linear analyses
is limited by the non-linear nature of the Einstein equations, and
in particular, the presence of perturbations is likely to affect
the evolution of the background cosmology at the non-linear level.
The back reaction of short wavelength gravity waves on an
expanding Friedmann-Robertson-Walker (FRW) cosmology is a
well-understood problem \cite{BHI}. However, in the context of
inflationary and post-inflationary cosmology, the scalar metric
fluctuations (fluctuations coupled to energy density and pressure
perturbations) are believed to dominate over the effects of
gravity waves, and thus, it is of great interest to understand the
possible back reaction of these scalar perturbations on the
cosmological background. Furthermore, during inflation, the phase
space of infrared modes (defined as modes with wavelength greater
than the Hubble radius) grows exponentially, whereas the phase
space of the ultraviolet modes does not grow and since the
amplitude of the associated metric fluctuations of these infrared
modes does {\it not} decrease in time , the back reaction of these
infrared modes may grow to be significant. There have been
different approaches to address this question. In \cite{MAB,ABM},
the effective energy-momentum tensor formalism of \cite{BHI} was
generalized to study the back reaction of infrared modes on the
spatially homogeneous component of the metric. The result was
that, in a slow-roll inflationary background, the back reaction of
infrared modes takes the form of a negative cosmological constant.
This was later confirmed in \cite{aw0} using very different
techniques.

The objections (see \cite{unruh}) to these analyses are mainly
concerned with the fact that \cite{BHI,MAB,ABM} do not consider a
local observable quantity, and even if the effect is not simply a
gauge artifact, it is still not clear how a local observer could
distinguish the back reaction of long-wavelength modes from an
initial condition ambiguity in a homogenous universe. More
importantly, calculating an ``observable'' from the spatially
averaged metric will not in general give the same result as
calculating the spatially averaged value of the observable. A
later work by Abramo and Woodard in \cite{aw2} tried to resolve
this problem by identifying a local physical variable which
describes expansion rate of the universe, then calculating the
back reaction of cosmological perturbations on this quantity. A
different approach was introduced in \cite{GB1}, where a simple
variable describing the local expansion rate was considered and
calculated up to second order in the metric fluctuations in a
model with a single inflation field \footnote{This method has been
recently used in \cite{rasanen} to investigate the back reaction
of UV modes in late Universe.}. In this approach, the leading
contribution of back reaction of infrared modes to this variable
was evaluated. When evaluated at a fixed value of the matter field
$\varphi$, the only physical clock available in this simple
system, the dependence of the expansion rate on the clock time
takes on exactly the same form as in an unperturbed background
\footnote{Back reaction of infrared modes for two scalar field
models where one of the fields is the inflaton and the other one
is the clock was investigated in \cite{back2}.}. Thus, the leading
infrared back reaction terms had no locally measurable effect in
this system. Similar conclusions were reached in \cite{aw3}.

However, the fact that during inflation the phase space of
infrared modes is growing in time, i.e. modes are exiting the
Hubble radius, leads us to the conjecture that there might be a
locally measurable effect just due to these crossings, and if we
include the corrections due this effect, the results for single
scalar field models may change. Since these modes are generated in
the quantum mechanical ultraviolet limit, handling them in general
is very difficult. Attempts, so far, in order to model the
evolution through writing a classical Langevin-like equation for
the coarse-grained scaler field, the so-called {\it stochastic
approach} \cite{starobinsky}, fail to include the inflaton and
metric fluctuations self-consistently \footnote{A similar method
has been used in \cite{yokoyama} in a different context to
investigate the behavior of weakly self-interacting scalar field
in a de Sitter back ground.}. In this work, for the first time, we
devise a self-consistent stochastic framework to study the
evolution of the coarse-grained inflaton. While the approach uses
the results of linear (metric+inflaton) perturbation theory for
the ultraviolet and transitional modes, it is rigorously extended
to the {\it non-linear} infrared regime, which allows a consistent
study of the evolution of fluctuations through their entire
history. We then apply this framework to the problem of back
reaction in inflationary cosmology, and investigate how/when the
non-linear effects may change properties/predictions of the
standard slow-roll inflation.

 The structure of the paper is as follows: In Section II,
we explain how the linear theory of perturbations can be
self-consistently generalized to the non-linear infrared regime,
including both inflaton and metric fluctuations. We also introduce
$\Theta$, the local expansion rate, and how it is related to
metric perturbations. In Sec. III, we show that the quantum
generation of fluctuations can be modelled as a stochastic term in
the equation of motion for $\Theta$, coarse-grained over local
Hubble patches. In Sec. IV, we apply this equation to a chaotic
inflationary scenario with a quadratic potential
$V(\varphi)={1\over2}m^2\varphi^2$ and evaluate the local
expansion rate with back reaction corrections due to infrared
modes. Sec. V contains the generalization of the result from Sec.
IV to any power-law potential. In Sec. VI, we investigate if back
reaction can ever significantly change the inflationary
predictions for our observable universe. Finally, Sec. VII
summarizes our results and concludes the paper. Throughout this
paper, we use the natural units, and set the Planck mass equal to
unity, i.e. $8\pi G = c= \hbar =1$.

\section{Generalized Bardeen parameter in the nonlinear regime}
Our analysis is based on the consistent quantization of the
linearized metric and matter perturbations in a classical
expanding background space-time. In this framework, it can be
shown that, in the linear regime, the Sasaki-Mukhanov parameter
$v$ \cite{Sasaki,VM1}, defined as \footnote{The cosmological scale
factor is denoted by $a(t)$, the Hubble expansion rate by $H$, and
the scalar matter field by $\varphi$.}:
\begin{equation}
 v \, = \, a[\delta\varphi+(\frac{\dot{\varphi}}{H})\phi] \, = \,  \frac{a\dot{\varphi}}{H}\zeta,
 \end{equation}
  obeys the equation of motion of a free scalar field, on scales much smaller than the Hubble radius.
  Here, $\zeta$ is the
Bardeen parameter \cite{Bardeen}:
  \begin{equation}  \zeta \, = \, \phi - \frac{H}{\dot{H}}(H\phi+\dot{\phi}) \, , \end{equation}
 where $\phi$ is the scalar metric perturbations in the longitudinal gauge, in a spatially flat universe:
 \begin{equation}
 ds^2 \, = \, (1+2\phi)dt^2 - a^2(t)(1-2\phi)\delta_{ij}dx^i dx^j \,  .
 \end{equation}

We use the following ansatz for our non-linear metric:
\begin{equation}
  ds^2 \, = \, e^{2\phi}dt^2-e^{-2\psi} \delta_{ij}dx^i dx^j, \end{equation}
   which is a generalization of a metric with linear scalar perturbations.
   This ansatz, though not a general solution, represents the most important sector of the
   metric,
   as any such metric can be continuously connected to a metric with linear scalar
   perturbations, which is the dominant sector if perturbations are generated during slow-roll inflation \cite{LLrev}.
 We consider the simplest model of matter with a single scalar field $\varphi$, the inflaton.
 We also use a perfect fluid model to describe the expansion of the universe.
 The velocity four-vector field $u^{\alpha}$ can be considered to be tangential
 to a family of world lines of a matter fluid in a general space-time.
 This four-vector is normalized such that
  \begin{equation}
  \label{norm} u^{\alpha}u_{\alpha} \, = \, 1 \, ,
  \end{equation}
 where $\alpha$ runs over the space-time indices. In terms of this four-vector, the local expansion rate $\Theta$ is defined by
  \begin{equation}
  \Theta \, \equiv \, u^{\alpha}_{\,;\alpha} \,=~ 3 H_{\rm local}
  ,
  \end{equation}
   where $H_{\rm local}$ is the local Hubble constant. It represents the local expansion rate
   of the tangential surfaces orthogonal to the fluid flow. It is easy to verify that,
   for our specific choice of gauge, $\Theta$ follows the following relation
    \footnote{For a more detailed discussion about calculation of this parameter see \cite{GB1,back2}.},
    \begin{equation} \label{thetapsi}
     \Theta=-3\psi^{\prime},
     \end{equation}
 where a prime denotes the derivative with respect to $\tau$, the proper time,
 and is related to the coordinate time via $d\tau=e^\phi d t$.
The Lagrangian and the energy-momentum tensor for the inflaton are
given by:
 \begin{eqnarray} {\mathcal{L}} = \frac{1}{2}
\partial^{\alpha}\varphi \partial_{\alpha}\varphi -  V(\varphi),
\\  T^{\mu}_{\nu} = \partial^{\mu}\varphi\partial_{\nu}\varphi -
\mathcal{L}\delta^{\mu}_{\nu}.
 \end{eqnarray}
It can be shown that for slow-roll inflation, in the
long-wavelength limit, $\Theta$ satisfies the following condition
\cite{GB1},
\begin{equation} \label{thetaV}
\Theta=\sqrt{3V(\varphi)}.
\end{equation}
In \cite{niayesh} it is demonstrated that, in these limits, the
generalized Bardeen parameter:
 \begin{equation} \label{defzeta}
\tilde{\zeta} ({\bf x}) \, \equiv  \, \psi - \frac{1}{2}  \int
\bigl(\frac{\partial  \log(g)}{\partial \varphi}\bigr)^{-1}
d\varphi,
\end{equation}
remains constant in the classical non-linear regime \cite
{salopek1}. Here, $g$ is a function of the inflaton field,
obtained by integrating the $G_{0i}$ Einstein constraint
equations, and satisfies
\begin{equation} \label{gpsi}
2 \psi^{\prime} \, =\,  g(\varphi); ~\varphi^\prime =
\frac{\partial g(\varphi)}{\partial \varphi}. \end{equation}

It is easy to see that for linear perturbations $\delta
\tilde{\zeta}(x)$ reduces to the usual Bardeen parameter $\zeta$,
which is known to remain constant for adiabatic super-Hubble modes
\footnote{In linear perturbation theory, $\delta\phi$ is forced to
be equal to $\delta\psi$ by the off-diagonal $G_{ij}$ equations
for all theories in which the off-diagonal components of $T_{ij}$
vanish to the linear order. This is in particular the case for
scalar field matter.}.

Combining the above equation, with Eqs. (\ref{thetapsi}) and (
\ref{defzeta}), we see that \begin{equation} \label{zetaprime}
\tilde{\zeta}^{\prime}=-{\Theta \over
3}-\frac{1}{2}(\frac{\partial \log(\Theta)}{\partial
\varphi})^{-1}\varphi^{\prime} = 0, \end{equation}  in the
classical long-wavelength limit. In the next section, we  show how
quantum fluctuation affect Eq. (\ref{zetaprime}).

\section{coarse-graining the expansion rate}

 In this section, we use coarse-graining on the scale of the
Hubble radius to provide a framework to analyze back reaction, and
obtain a new derivation of the equation of motion for stochastic
inflation \cite{Starob}, which takes into account the metric
fluctuations.

Let us define the coarse-grained generalized Bardeen parameter,
$\zeta_c$, in the Fourier space, as:
\begin{equation}\label{cg}
\tilde{\zeta}_{c}({\bf k})=\theta(aH-k)\tilde{\zeta}({\bf k})
\end{equation}
where $\theta$ is the step function (not to be mistaken with
$\Theta$, the local expansion rate) which is zero for $k > aH$,
and is one for $k\leq aH$ \footnote{This is the simplest example
of a cutoff function. In general, one can use any function which
is close to one for $k < aH$ and tends to zero as $k$ goes to
infinity fast enough to eliminate the ultraviolet divergence.}.
Taking the time derivative of the above equation; the time
derivative of $\tilde{\zeta}$ vanishes for infrared modes and the
time derivative of $\theta$ is a delta function at $k=aH$, which
yields

\begin{equation}
\tilde{\zeta}^\prime_c({\bf k})=\tilde{\zeta}({\bf k})(H+{H^\prime
\over H})\delta(1-{k\over aH}).
\end{equation}
This simply implies that $\tilde{\zeta}^\prime_c(x)$ in real space
is related to $\tilde{\zeta}(k)$ in the following way
\begin{equation}\label{xk}
\tilde{\zeta}^\prime_c({\bf x})=(H+{H^\prime \over H})\int {{\bf
d^3 k} \over (2\pi)^3} e^{i{\bf k.x}}\tilde{\zeta}({\bf
k})\delta(1-{k\over aH}).
\end{equation}
Two-point function of $\tilde{\zeta}(k)$ can be described in terms
of its power spectrum, $P_{\tilde{\zeta}}(k)$, defined by:
\begin{equation}
\langle\tilde{\zeta}({\bf k_1})\tilde{\zeta}({\bf
k_2})\rangle=(2\pi)^3\delta({\bf k_1-k_2})P_{\tilde{\zeta}}(k_1).
\end{equation}
 Assuming that Hubble-scale perturbations are linear, $P_{\tilde{\zeta}}(k)$ can be derived in the linear theory of
quantum generation of perturbations \cite{MFB},
\begin{equation} \label{pwrspc}
P_{\tilde{\zeta}}(k) \simeq P_{\zeta}(k) = {H^4 \over 2 k^3
\dot{\varphi}^2}|_{k=aH}.
\end{equation}
Using this knowledge for $\tilde{\zeta}(k)$ and assuming
${H^\prime \ll H^2}$ (the slow-roll condition) in the above
equation, we can calculate the two-point function for
$\tilde{\zeta}^\prime_c({\bf x},\tau)$. Note that $H$ in Eqs.
(\ref{cg}-\ref{pwrspc}) is the local Hubble constant, and can be
replaced by $\Theta /3$. Combining Eqs. (\ref{xk}-\ref{pwrspc}),
after performing few simple integrations, we arrive at:
\begin{eqnarray} &\,& \label{twopointzeta}
\langle\tilde{\zeta}^\prime_c({\bf
x_1},\tau_1)\tilde{\zeta}^\prime_c({\bf
x_2},\tau_2)\rangle\\&=&-{1\over 648\pi^2}{\Theta^5\over
\Theta^\prime}\delta(\tau_1-\tau_2)~{\rm sinc}(aH{\bf |x_1-x_2|})
\nonumber
\end{eqnarray}
where ${\rm sinc}(x)={\sin(x) \over x}$ and ${\rm sinc}(0)=1$. Let
us express $\tilde{\zeta}^\prime_c$ as
\begin{equation} \label{defxi}
\tilde{\zeta}^\prime_c(x,\tau)\equiv{1\over
36\pi}\bigl({-\Theta^5\over 2
\Theta^\prime}\bigr)^{1/2}\xi(x,\tau),
\end{equation}
where the variable $\xi$ is a random Gaussian field with:
\begin{equation}\label{xi2} <\xi({\mathbf{x_1}},t_1)
\xi({\mathbf{x_2}},t_2)> \, = \, \delta(\tau_1-\tau_2)~{\rm
sinc}(aH{\bf |x_1-x_2|}) . \end{equation}

Now, plugging the coarse grained value of $\tilde{\zeta}^\prime$
from Eq. (\ref{defxi}) into the left hand side of Eq.
(\ref{zetaprime}), we may obtain the coarse-grained field
equation:
\begin{equation}
\label{maineqn} \varphi^{\prime} \, = \, \frac{\partial
\log(\Theta)}{\partial \varphi}\bigl[-{2 \over 3} \Theta
- {\sqrt{2}\over 36\pi}({-\Theta^5\over \Theta^\prime})^{1\over2}\xi(x,t)\bigr] \, . \end{equation}

\section{The case of a quadratic potential}

In this section, we apply the coarse-graining method, developed in
the previous section, to a chaotic inflationary scenario with a
quadratic potential, and investigate the evolution of the
background during inflation under the effect of second-order back
reaction.

Here, we assume that the potential for the inflaton field,
$\varphi$, has a quadratic form: \begin{equation}
V(\varphi)={1\over2}m^2\varphi^2
\end{equation} Plugging this into Eq. (\ref{thetaV}), and
then (\ref{maineqn}),  the equation of motion for $\Theta$ reduces
to
\begin{equation}\label{eqnquad} \Theta^\prime+m^2={\sqrt{2}\over 24\pi}m\Theta^{3\over2}\xi(x,\tau).
\end{equation}
This equation can be solved perturbatively, taking $\xi$ as the
perturbation variable. $\Theta$ can be approximated by
\begin{equation} \Theta=\Theta_0+\Theta_1+\Theta_2,
\end{equation}
where $\Theta_0$, $\Theta_1$ and $\Theta_2$ are zeroth, first and
second order in $\xi$, respectively. As the first step in solving
Eq. (\ref{eqnquad}), we note that $\Theta_0$ is simply the
expansion rate in the absence of perturbations, \begin{equation}
\Theta_0=-m^2\tau,\\ \end{equation} where our convention for
$\tau$ is so that it has a negative initial value $\tau_{in}$ and
goes to zero towards the end of inflation, so that $\Theta$
decreases as inflation proceeds. Now we substitute $\Theta_0$ back
into Eq. (\ref{eqnquad}) to find $\Theta_{1}$,
\begin{equation} \Theta_1={\sqrt{2} \over 24\pi} m
\int(-m^2\tau)^{3/2}\xi({\bf x},\tau) d\tau \end{equation} This
step is easily followed by calculation of $\Theta_2$.
\begin{equation} \label{theta2} \Theta_2={m^6\over 288\pi^2}
\int_{\tau_{in}}^{\tau}\int_{\tau_{in}}^{\tau_1}d\tau_1d\tau_2(\tau_{_1}\tau_{_2}^3)^{1/2}\xi({\bf
x},\tau_1)\xi({\bf x},\tau_2).
\end{equation}
Since $\xi$ is a random Gaussian variable, we are not interested
in the spontaneous value of $\Theta$, and instead investigate the
statistical behavior of $\langle\Theta\rangle$, the expectation
value of the expansion rate over time as more infrared modes cross
the hubble radius and may build up a back reaction effect. As
$\Theta_0$ does not have a $\xi$ term, and the expectation value
of $\Theta_1$ vanishes over time, we have:
\begin{equation}
\langle\Theta\rangle=\Theta_0+\langle\Theta_2\rangle,
\end{equation}
where $\langle\Theta_2\rangle$ can be derived from Eqs.
(\ref{theta2}) and (\ref{xi2}), which finally enables us to
calculate expansion rate with corrections due to stochastic back
reaction:
\begin{equation} \langle\Theta\rangle=-m^2\tau+{m^6\over1728\pi^2}(\tau^3-\tau_{in}^3).\end{equation}

Contrary to our expectation from \cite{ABM}, we can see that the
back reaction term has a positive value and it is increasing with
time, as if it is speeding up inflation. However, as we see in the
next section, this is merely a result of our choice to average
over hyper-surfaces of constant {\it proper} time, while other
choices of time hyper-surfaces may lead to different forms, or
even a different sign, for the back reaction term.

Nevertheless, as both $\Theta$ and $\tau$ are observable, we may
conclude that the back reaction {\it does not vanish} during
inflation and since its magnitude is increasing with time, it may
become important if inflation lasts long enough. Notice that this
correction is, in fact, due to wavelengths larger than the Hubble
radius as, using Eq. (\ref{theta2}), $\langle\Theta_2\rangle$
could be expressed as an integral over $k$, the wave number
instead of $\tau$, which would yield
\begin{equation} \langle\Theta_2\rangle={1\over 594\pi^2}\int d\ln
k \cdot k^3
P_{\zeta}(k) \cdot {\Theta_0^{\prime 2}\over\Theta_0^3}\bigl|_k,\\
\end{equation} and the above integral is only over modes which
have already crossed the Hubble radius. In this sense, this is an
infrared back reaction effect.

\section{Generalization to power law
potentials}

In this section, we generalize the calculations of the last
section to power law inflaton potentials. Also, rather than proper
time, we use the inflationary e-folding number as our time
variable, as it has more direct and intuitive observational
relevance. The number of e-foldings since the beginning of
inflation is
\begin{equation}
N_e(\tau)={1\over3}\int_{\tau_{in}}^{\tau}\Theta d\tau.
\end{equation}

Using Eq. (\ref{thetapsi}), we can see that $\psi$ may be used as
a measure of e-folding:
\begin{equation}\label{net}
N_e(\tau)=\psi_{in}-\psi(\tau),
\end{equation}
and thus, in order to find how $\Theta$ changes as a function of
$N_e$, it is sufficient to express it in terms of $\psi$.
Combining Eq. (\ref{twopointzeta}) with ${d\over
d\tau}=\psi^{\prime}{d\over d\psi}$, we can re-write Eq.
(\ref{maineqn}) as
\begin{equation} \label{maineqn2}
{d\varphi \over d\psi} \, = \, \frac{\partial
\log(\Theta)}{\partial \varphi}\bigl[2- {\sqrt{2}\over
6\pi}({\Theta^3\over d\Theta/ d\psi})^{1\over2}\xi({\bf
x},\psi)\bigr],
\end{equation}
where $\xi$ in the above equation like is, again, a random
Gaussian field, defined by
\begin{equation}\label{xit2} <\xi({\mathbf{x_1}},\psi_1)
\xi({\mathbf{x_2}},\psi_2)> \, = \, \delta(\psi_1-\psi_2)~{\rm
sinc}(aH{\bf |x_1-x_2|}) . \end{equation}

 In the case of single
scalar field inflationary models with a power law potential
\begin{equation}
V(\varphi)=M^4\varphi^\alpha ,
\end{equation}
using the procedure described in the previous section, Eq.
(\ref{maineqn2}) can be solved to obtain $\Theta$ in terms of
$\psi$:
\begin{eqnarray}\label{backp}
\langle\Theta\rangle&&=\alpha^{\alpha/4}(3M^4)^{1/2}\psi^{\alpha/4} +\nonumber\\
&&{ (3M^4)^{3/2}\alpha^{1+3\alpha/4} \over
288\pi^2(\alpha+4)}\psi^{{\alpha\over4}
-2}(\psi^{2+\alpha/2}-\psi_{in}^{2+\alpha/2}).
\end{eqnarray}

Note that we assumed $\psi \rightarrow 0$ towards the end of
inflation. The result may be simplified for the case of a
quadratic potential:
\begin{eqnarray}\label{thet}
\langle\Theta\rangle&=&\sqrt{2}(3M^4)^{1/2}\psi^{1/2}\nonumber\\
&+&{ (3M^4)^{3/2}\sqrt{2} \over 432\pi^2}\psi^{
-3/2}(\psi^{3}-\psi_{in}^{3}).
\end{eqnarray}

We see that, if averaged over the hyper-surfaces of constant
e-folding rather than constant proper time, the second order back
reaction has a negative sign, i.e. it slows down inflation. We
again believe that this is a real effect, as both $\Theta$ and
$N_e$ are observable parameters.

\section{Does back reaction change inflationary predictions?}

Let us calculate the ratio of the second order term in Eq.
(\ref{backp}) to the zeroth order term
\begin{equation}\label{ratio}
{\langle\Theta_2\rangle \over \Theta_0}\simeq{3M^4 \over
288\pi^2}\alpha^{1+\alpha/2}\psi_{in}^{2+\alpha/2}\psi^{-2},
\end{equation}
 if $\psi \ll \psi_{in}$, close to the end of inflation.
  Currently, the most stringent constraints on the inflationary
potential is due to the observations of the cosmic microwave
background anisotropies by the WMAP satellite \cite{hiranya}. The
observed amplitude of anisotropies at the observable horizon scale
require:
\begin{eqnarray}
\Delta^2_{\cal R} \equiv \frac{k^3 P_{\zeta}(k)}{2\pi^2} &=&
\frac{V^3(\varphi)}{12\pi^2 V^{\prime 2}(\varphi)} = (2.1 \pm 0.3)
\times 10^{-9}, \nonumber\\
{\rm for}~~ k &=& 0.002 ~{\rm Mpc}^{-1},
\end{eqnarray}
with $V^{\prime}$ being the derivative of $V$ with respect to
$\varphi$.
   Then we can use the simple slow-roll prediction
   (first term in Eq. \ref{thet}, along with Eq. \ref{thetaV})
  to relate $\varphi$ to $\psi$, and express $\Delta^2_{\cal R}$
  in terms of $\psi$:
 \begin{equation}
 \Delta^2_{\cal R} = \frac{M^4}{12 \pi^2} \alpha^{\alpha/2 -1}
 \psi^{1+\alpha/2}.
 \end{equation}

   Assuming that $\psi_{in} \simeq N_e$ (Eq. \ref{net}), and that inflation ends
   when $V^\prime \sim V$ (in Planck units), Eq. (\ref{ratio})
   can be expressed in terms of $\Delta_{\cal R}$:
\begin{equation}
{\langle\Theta_2\rangle \over \Theta_0} \simeq \frac{1}{8}
\Delta^2_{\cal R} N_e = 2 \times 10^{-8}
\left(\frac{\Delta^2_{\cal R}}{2 \times 10^{-9}}\right)
\left(\frac{N_e}{70}\right).
\end{equation}

   Although, this result is specific to the power law inflaton potentials,
  we expect a similar order of magnitude for any smooth potential
  that is consistent with observational bounds.
\begin{figure}
       \includegraphics[angle=-90, width=\linewidth]{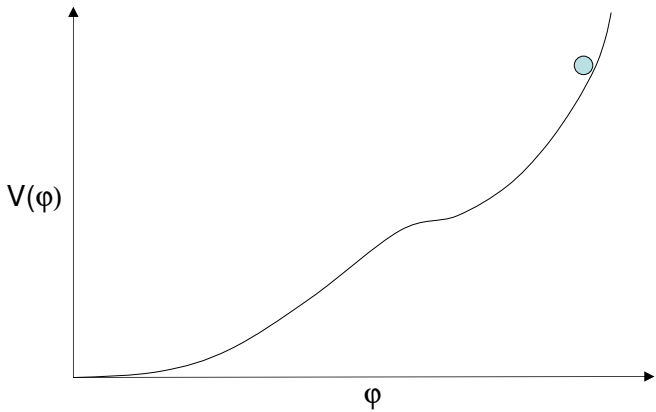}
      \caption{\label{signal} The example of an inflaton potential
      with a flat portion. Back reaction can dominate the slow-roll inflation in
      this portion of the potential.}
  \end{figure}
  Does this mean that back reaction will always be negligible
  during inflation? Of course, the number of e-foldings, $N_e$, could be
  arbitrarily large. Indeed, after enough e-foldings, we end up with the
  self-similar chaotic inflationary picture which results from the
  stochastic formulation \cite{linde}. However, only $\sim 70$ of these e-foldings
  could fit in our local Hubble patch today, and any effect on the power spectrum due to the back reaction of
  the super-Hubble modes is inaccessible to observations.


  Another possibility is that the inflaton potential may not be smooth.
  The observational bounds on $\Delta^2_{\cal R}$ hold from the
  observable horizon ($\sim 10^4 ~{\rm Mpc}$) down to the scale of the Lyman-$\alpha$ forest ($\sim 1 ~{\rm
  Mpc}$) \cite{croft}, constraining only about 10 e-foldings in the inflaton potential. An unusually
  flat portion in the inflaton potential (see Fig. \ref{signal}) may yield a
  significantly larger amplitude of fluctuations, e.g.
  $\Delta^2_{\cal R}\sim 0.1$, and be inaccessible to cosmological
  observations if it happens at small enough scales. However, such
  a large amplitude may result in the production of Primordial Black
  Holes (PBHs), as the modes enter the Hubble radius, which
  results in further constraints on $\Delta_{\cal R}$.

  To study the production of PBHs \cite{green}, let us look at the comoving scale of $R \lesssim 100 ~{\rm
  kpc}$, with the primordial amplitude of fluctuations $\Delta_{\cal R}$.
   This scale enters the Hubble radius when $aH \sim \pi/R$, which happens at redshift $z$,
  roughly given by
  \begin{equation}
  z \sim \frac{\pi \sqrt{z_{eq}}}{H_0 R \sqrt{\Omega_m}} \sim
  10^9 R^{-1}({\rm kpc}),
  \end{equation}
  where $H_0 \simeq 70 ~{\rm km/s/Mpc}$ and $\Omega_m \simeq 0.3$ are today's Hubble constant and
  density parameter respectively, while $z_{eq} \sim 3\times 10^3$ is the
  redshift of radiation-matter equality. As the modes of scale $R$
  enter the Hubble radius, a fraction, $f$, of the Hubble patches
  have a fluctuation amplitude larger than $1$, and thus may form
  PBHs, where $f$ is given by
  \begin{equation}\label{ff}
  f(\Delta_{\cal R}) \sim \int^{\infty}_1 \exp[-\frac{\zeta^2}{2\Delta^2_{\cal R}}]
  \frac{d\zeta}{\sqrt{2\pi \Delta^2_{\cal R}}}  \sim \Delta_{\cal R} \exp[-\Delta^{-2}_{\cal R}/2].
  \end{equation}

    The current density in PBHs in units of the critical density,
    $\Omega_{\rm PBH}$, is given by
   \begin{equation}\label{om}
    \Omega_{\rm PBH} \simeq f(\Delta_{\cal R}) (\frac{z}{z_{eq}})
    \epsilon,
   \end{equation}
    where $10^{-4} < \epsilon <1$, is the fraction of the mass in a Hubble patch
    that goes into the PBH \cite{hawke}. Requiring that
    $\Omega_{\rm PBH} < \Omega_m$ yields
    \begin{equation}\label{omega}
    f(\Delta_{\cal R}) < \frac{z_{eq}\Omega_m}{z \epsilon}
    \lesssim 10^{-2} R({\rm kpc}).
    \end{equation}
    For $\Delta^2_{\cal R} = 0.1$, Eq. (\ref{ff}) gives $f(\Delta_{\cal R})
    \sim 10^{-3}$, and thus Eq. (\ref{omega}), in combination with large scale structure observations, require
    \begin{equation}
       0.1 ~ {\rm kpc} \lesssim  R \lesssim 100 ~ {\rm kpc}.
     \end{equation}

     Other astrophysical observations may put tighter constraints on a population of PBHs in the universe \cite{PBHcons}.
     Instead of going through these constraints, let us give an example. Assume at $R
     = 10 ~{\rm kpc}$, $\Delta^2_{\cal R} = 0.1$ and $\epsilon = 10^{-4}$.
     Using Eq. (\ref{om}), $\Omega_{\rm PBH} \sim 0.01 \Omega_m$,
     while the PBH mass, $M_{\rm PBH}$ is $\epsilon M_H$, with
     $M_H$, being the Hubble mass at the Hubble entry:
     \begin{equation}
     M_H \sim 10^{17} M_{\odot} \left(\frac{z_{eq}}{z}\right)^2 \sim (10^5~M_{\odot}) R^2({\rm
     kpc}).
     \end{equation}
     Therefore, with our choice of parameters, about $1\%$ of the
     mass of the universe will end up in $\sim 10^3 M_{\odot}$
     black holes. This is consistent with current
     astrophysical constraints on such objects \cite{PBHcons}.

      To summarize this section, we showed that if we assume a smooth inflaton
      potential, in single matter field models infrared back reaction (during the last 70 e-foldings of inflation) should be negligible, at
      least to the 2nd order (one loop) in the amplitude of
      perturbations. However, if we relax this assumption, it is possible to construct
      an inflaton potential that allows a significant
      back reaction effect during the slow-roll inflationary
      period, without breaking any observational
      constraints.

\section{Conclusions}

  We use a stochastic coarse-graining method to approach the
  problem of infrared back reaction in a single scalar field inflationary cosmology.
  Using the coarse-grained generalized Bardeen parameter, we
  self-consistently derive the stochastic Langevin equation that governs the evolution of
  the local expansion rate. Solving this equation to second order (one loop) for chaotic
  inflation with a power law potential, we see that there is a non-vanishing
  average back reaction term due to the infrared modes that have
  left the Hubble radius. Although, the size of this effect will
  be negligible (within the observable universe) for smooth inflaton potentials, we show that it is
  possible to have non-smooth potentials which allow significant
  back reaction happening on sub-horizon scales, without breaking any observational
  constraints. A possible implication of such non-smooth potentials is the generation of a population of massive primordial black holes,
  which could be further used to constrain or verify these models.

\medskip
\centerline{Acknowledgments}
\medskip

We would like to thank Robert Brandenberger for many useful
insights during the course of this research. G.G. would like to
thank Brown University for a Dissertation Fellowship which this
research was in part supported with.

\end{document}